\documentclass[aip,jap,reprint]{revtex4-1}
\pdfoutput=1

\usepackage[usenames,dvipsnames]{color}
\usepackage{graphicx}
\usepackage{tabularx}
\usepackage[T1]{fontenc}
\usepackage[cp1250]{inputenc}
\usepackage{dcolumn}
\usepackage{amsmath,amsfonts,amssymb}
\usepackage{multirow}
\usepackage[rightcaption]{sidecap}
\usepackage{float}
\newcommand{\CdMnTe}{Cd$_{1-x}$Mn$_x$Te}

\newcommand{\ZnMgTe}{Zn$_{1-x}$Mg$_x$Te}

\newcommand{\beq}{\begin{equation}}
\newcommand{\eeq}{\end{equation}}

\begin{document}

\title{Engineering the hole confinement for CdTe-based quantum dot molecules}

\author{Ł.~Kłopotowski}
\email[Corresponding author: ]{lukasz.klopotowski@ifpan.edu.pl}
\affiliation{Institute of Physics, Polish Academy of Sciences, Al. Lotników 32/46, 02-668 Warsaw, Poland}

\author{P.~Wojnar}          \affiliation{Institute of Physics, Polish Academy of Sciences, Al. Lotników 32/46, 02-668 Warsaw, Poland}

\author{S.~Kret}            \affiliation{Institute of Physics, Polish Academy of Sciences, Al. Lotników 32/46, 02-668 Warsaw, Poland}

\author{M.~Parlińska-Wojtan}            \affiliation{Facility for Electron Microscopy and Sample Preparation, Center for Microelectronics and Nanotechnology, Faculty of Mathematics and Natural Sciences, University of Rzeszów, ul. Pigonia 1, 35-959 Rzeszów, Poland}

\author{K.~Fronc}           \affiliation{Institute of Physics, Polish Academy of Sciences, Al. Lotników 32/46, 02-668 Warsaw, Poland}

\author{G.~Karczewski}      \affiliation{Institute of Physics, Polish Academy of Sciences, Al. Lotników 32/46, 02-668 Warsaw, Poland}

\author{T.~Wojtowicz}       \affiliation{Institute of Physics, Polish Academy of Sciences, Al. Lotników 32/46, 02-668 Warsaw, Poland}

\date{\today}

\begin{abstract}

We demonstrate an efficient method to engineer the quantum confinement in a system of two quantum dots grown in a vertical stack. We achieve this by using materials with a different lattice constant for the growth of the outer and inner barriers. We monitor the resulting dot morphology with transmission electron microscopy studies and correlate the results with ensemble quantum dot photoluminescence. Furthermore, we embed the double quantum dots into diode structures and study photoluminescence as a function of bias voltage. We show that in properly engineered structures, it is possible to achieve a resonance of the hole states by tuning the energy levels with electric field. At the resonance, we observe signatures of a formation of a molecular state, hybridized over the two dots.

\end{abstract}


\maketitle

\section{Introduction}

Manipulation of a quantum state lies at the heart of quantum information processing. There is, thus, a need for systems where external factors, such as gate voltage or magnetic field, can be used to tailor the quantum properties. One of such systems are coupled semiconductor quantum dots (QDs) -- quantum dot molecules (QDMs)\cite{led96,sch97} -- in which tunnel coupling between atomic-like states leads to hybridization of wave functions into molecular-like orbitals. QDMs were proposed as promising candidates for entangling of quantum states and performing operations on quantum logic gates. \cite{bru97,bur00,bay01} The first step on the path to realization of these concepts is obtaining a coherent and tunable coupling between the single dot qubits. There are two basic requirements: (i) the QDs have to reside at a distance allowing for efficient tunneling (i.e. a few nanometers) and (ii) the energy levels have to be close enough for an external field to bring them to resonance. The spatial correlation can be achieved for both vertically \cite{xie95} and laterally \cite{wan08} coupled QDMs. In the former case, the nucleation of dots in the top layer occurs as a result of a strain field emerging from the bottom one. In the latter, a patterning technique is employed to directly position two QDs next to one another in a single layer. For vertically stacked QDs, the top one is usually larger owing to a smaller amount of elastic energy present during its formation.\cite{xie95} In order to obtain QDs with a similar morphology and thus having energy levels close to resonance an in the the case of InAs QDs, indium flush technique is usually employed.\cite{was99} Photoluminescence (PL) experiments are used to reveal the coupling in QDMs and, in particular, to show that the coupling can be tuned with an electric field.\cite{kre05,ort05,sti06,bra06,sch07prb,dot10} The coherent coupling manifests itself as an anticrossing between the interdot and intradot exciton transitions. The former exciton is formed from an electron and a hole occupying adjacent dots, while in the latter both carriers are confined to the same dot. The anticrossing appears when the single carrier states in the neighboring QDs are brought to resonance, where the orbital wave functions hybridize into bonding and antibonding states. The QDMs were further employed, e.g., to demonstrate a coherent coupling of a two-spin qubit \cite{kim11}, to store an electron-hole pair for a time exceeding its lifetime by three orders of magnitude,\cite{boy11} or to employ resonant fluorescence to read-out the qubit spin state.\cite{vam10}

Most of the achievements in this field were accomplished on InGaAs-based nanostructures, where fabrication methods are well established. On the other hand, CdTe-based QDMs offer two important advantages. The first one is related to relatively low abundance of non-zero spin isotopes (25\% of Cd and 8\% of Te) and the fact that the nuclear spin of these isotopes is only $I=1/2$, compared to InGaAs, where 100\% In atoms carry $I=9/2$ and 100\% of Ga and As atoms carry $I=3/2$. As a result, the carrier-nuclei hyperfine interaction leading to qubit decoherence is expected to be much weaker in CdTe QDs. The second advantage is the possibility of doping the QDs with single \cite{mai06} and multiple \cite{mak00} transition metal ions. In the latter case of semimagnetic dots, the carrier-ion exchange interaction results in $g$-factors that can be more than two orders of magnitude larger than in intrinsic CdTe QDs. This opens up a whole new field of possibilities including using the individual dopant spins as carriers of information \cite{bes04,gor09,kob14} or employing semimagnetic QDs as optically programable nanomagnets \cite{mac04,fer04,klo11pol}. In the context of QDMs, the presence of magnetic dopants should allow for an alternative method for tuning of the inter-dot coupling: with the magnetic field via the giant Zeeman effect.\cite{bac08} Moreover, combining the electric and magnetic fields in these semimagnetic structures is predicted to allow fast spin qubit rotations.\cite{lya12}

In this work, we demonstrate how to engineer the confinement in two CdTe QDs grown in a vertical stack. We start with showing that, similarly to the InGaAs system, the dots in adjacent layers indeed grow in vertical stacks. Also, analogously to the III-V system, if the dots are grown in a uniform matrix, the top ones are significantly larger. This leads to a large energy detuning for states in the neighboring dots, which in turn results in an inability to bring the carrier states to resonance with external electric field. We resolve the problem of large energy detuning by applying a method of strain compensation. We grow the bottom QD on a \ZnMgTe\ barrier and then {\em increase} the amount of elastic energy for the formation of the top dot by growing a ZnTe spacer layer. We monitor the resulting QD morphology with transmission electron microscopy in correlation with ensemble QD PL. Finally, we show that the QDMs with engineered confinement indeed exhibit anticrossings characteristic of the molecular-like coupling. With a perspective of applying the method to semimagnetic QDMs, we aim at obtaining hybridization of holes states, since the hole $g$-factors in semimagnetic dots are roughly 4 times larger than for electrons.

\section{Samples and Experiment}

The QDs are fabricated using a modified Stranski-Krastanow procedure, well established for the formation of CdTe dots.\cite{tin03,woj10}. In this method, a layer of CdTe is deposited on the barrier material possessing a smaller lattice constant. The accumulated elastic energy is relaxed at an expense of surface energy of creating facets. The relaxation is catalyzed by covering the strained CdTe layer with amorphous tellurium and its subsequent desorption at a higher temperature. All the dots used in these studies are formed from a 6 monolayer thick CdTe layer. In the first series of samples (S1), the two layers of dots are formed between ZnTe barrier layers and separated with a ZnTe spacer layer, grown by previously calibrated atomic layer epitaxy method. In the second series (S2), in order to control the confinement (see below), the outer barriers are made of ternary Zn$_{0.88}$Mg$_{0.12}$Te, while the dots are separated, as in the first series, with ZnTe. In S1, we investigate four samples with the width of the ZnTe layer, determined from the TEM studies, equal to 1.2, 2.5, 3.8, and 5 nm. In S2, we study samples with this width equal to 5 and 7 nm.

For the experiments involving tuning of the hole levels in the neighboring dots, the S1 and S2 structure design modified via doping to enable application of bias. In the case of S1, the dots are grown in the intrinsic region of {\em p-i-}Schottky structures. In the case of S2, they are grown in the intrinsic region of {\em n-i-p} diodes. The former and latter structures allow to apply an electric field antiparallel and parallel to the growth axis, respectively. As discussed below, these field directions are necessary to bring the hole states to resonance.\cite{bra06} {\em n-} and {\em p-}type doping is obtained by incorporation of iodine and nitrogen dopants, respectively.\cite{klo11,klo14}

Ensemble PL studies are performed with a 405 nm solid-state laser used as the excitation source. The samples are placed in closed-circle cryostat and the measurements are performed at 5 K. For single QD studies, the samples are cooled down in a cold-finger cryostat to 10 K. A 532 nm solid-state laser beam is focused onto a 2 $\mu$m spot with a microscope objective. In order to limit the number of excited QDs, on top of the samples we deposit a metallic shadow mask with apertures of 200 nm in diameter. This metal layer also forms the top electrical contact. The bottom contact is established after wet chemical etching, down to a thick, doped bottom layer, {\em p-} or {\em n-}type for S1 and S2 samples, respectively.\cite{klo14}

The QD morphology is analyzed by transmission electron microscopy (TEM). The samples are glued with a film slide and mechanically polished using diamond laps with different grit sizes. The obtained foils are thinned down to electron transparency by Ar+ ion milling or focused Ga ion beam milling. The cross-sections of the samples are analyzed in the scanning transmission electron microscopy (STEM) mode using the high angular annular dark field (HAADF) detector operating at 200-300 kV. Energy dispersive X-ray spectroscopy (EDX) analysis is performed by the in-column EDX system. The local strain component between the [002] planes is determined by geometric phase analysis of the STEM images\cite{hyt98} and evaluated relative to the crystallographic matrix [002] lattice spacing.

\section{Results and Discussion}

\begin{figure}[!h]
  \includegraphics[angle=0,width=.50\textwidth]{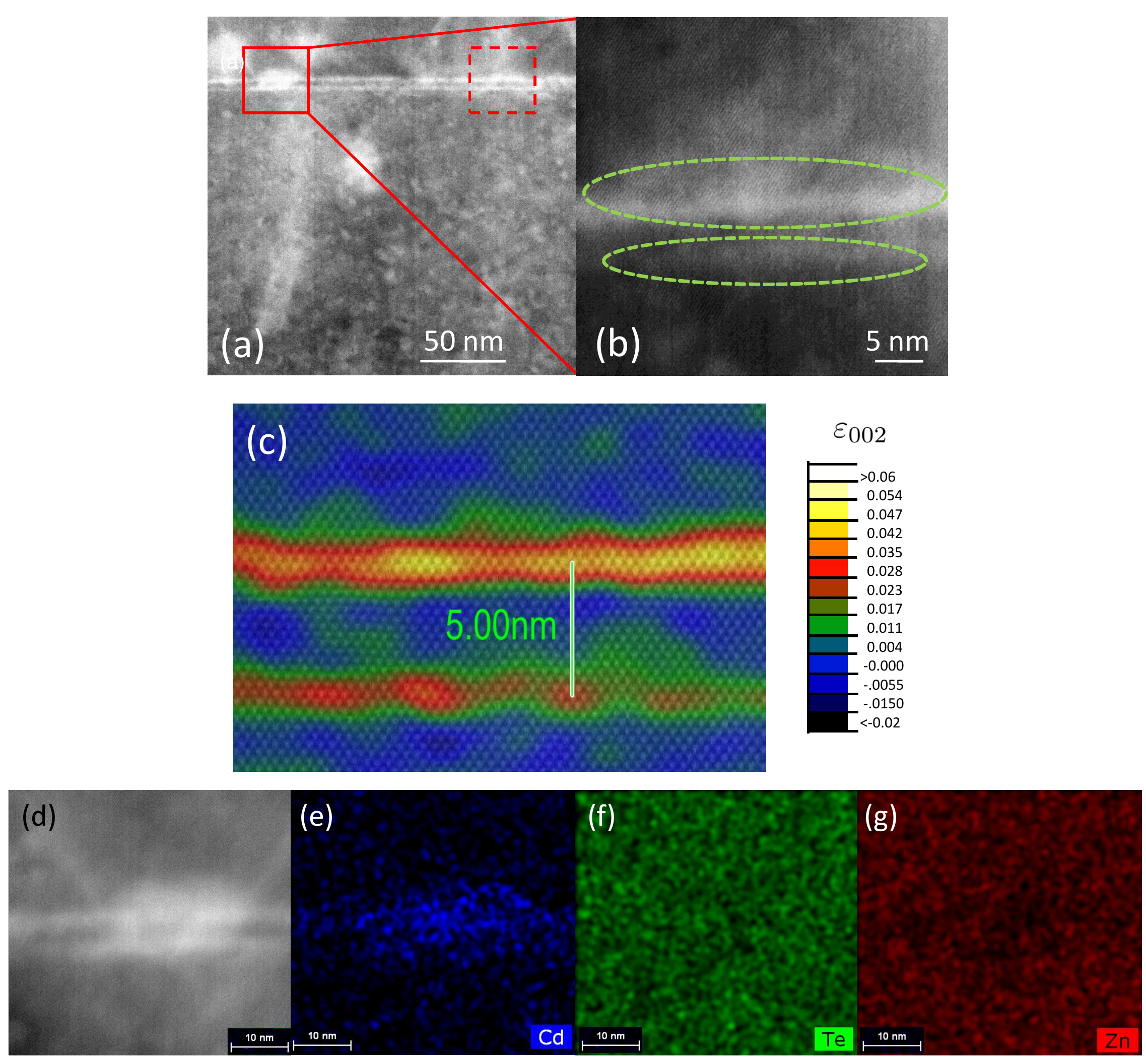}
  \caption{STEM analysis of S1 sample with the width of the ZnTe spacer of 5 nm. (a) HAADF image showing two layers of QD. A vertical stack is seen in the solid square and a single dot in the top layer in the dashed square. (b) Atomic resolution close-up of the solid square region showing the two stacked QDs. (c) Color coded, atomic resolution lattice distortion map pointing to a larger incorporation of cadmium atoms into the top QD layer.(d)-(g) EDX analysis of a stacked pair of QDs. (d) STEM HAADF image of a pair of quantum dots with the corresponding EDX maps of (b) Cd; (c) Te; (d) Zn.}
  \label{zntem}
\end{figure}

We start the discussion of experimental results with the analysis of transmission electron micrographs obtained on the S1 samples -- with the two layers of QDs grown in a ZnTe matrix. In Fig. \ref{zntem}(a), we show a representative HAADF image. The light spots correspond to regions rich in heavier atoms exhibiting also lattice distortions. The two layers of dots are clearly discerned. The spacer width for this sample is equal to 5 nm. On the left side of the image (inside the solid square), two dots grown in a vertical stack are seen. On the right (inside a dashed square), a single dot in a top layer is seen. We therefore conclude that the vertical spatial correlation of the dots in two layers is not perfect. However, from a survey of a few tens of dots, we estimate that the probability of finding a bottom QD aligned in a stack with a top QD is about 50\%. In some cases, a slight lateral shift amounting to less than half the QD diameter is found.

A close-up showing the vertical stack of QDs is presented in Fig. \ref{zntem}(b). From this atomic resolution image, the QD morphology can be assessed. From our survey, we infer that the dots exhibit a lens-like shape, with an average lateral size ranging from 10 to 30 nm, and height of about 2-3 nm, consistent with previous findings based on atomic force microscopy of similar samples.\cite{woj08} More importantly, Fig. \ref{zntem}(b) suggests that the top QD is richer in cadmium than the bottom one. This finding is confirmed by the analysis of lattice distortion along the growth axis. In Fig. \ref{zntem}(c), we show an atomic resolution map of local lattice distortion $\varepsilon_{002}$ evaluated as $\varepsilon_{002} = d/d_M-1$, where $d$ is the local lattice distance and $d_M$ is reference lattice distance of the surrounding matrix, i.e., ZnTe in this case. The map reveals that the top QD is taller than the bottom one. Moreover, we can evaluate average lattice distortions of about 0.02 and 0.05 for the bottom and top QDs, respectively. We interpret the positive values of $\varepsilon_{002}$ as an elongation of the elementary cells along the growth axis, resulting from a biaxial compressive strain in the QD plane. Further confirmation for increased amount of cadmium in the top layer with respect to the bottom one is found in the EDX maps shown in Fig. \ref{zntem}(d)-(g).

From the point of view of spectroscopic studies of inter-dot coupling, it is important to assess the relative confinement depths of the QDs in adjacent layers. In a CdTe/ZnTe heterostructure the valence band offset is nearly zero and thus the hole confinement is predominantly driven by strain. Since the lattice distortion parameter in the two layers differs by a factor of about 2.5, we expect a large energy difference between the energies of the hole states confined in the two QDs. On the other hand, the electrons are less affected, since the confinement in the conduction band is mainly due to the huge conduction band offset amounting to about 500 meV.

\begin{figure}[!h]
  \includegraphics[angle=0,width=.50\textwidth]{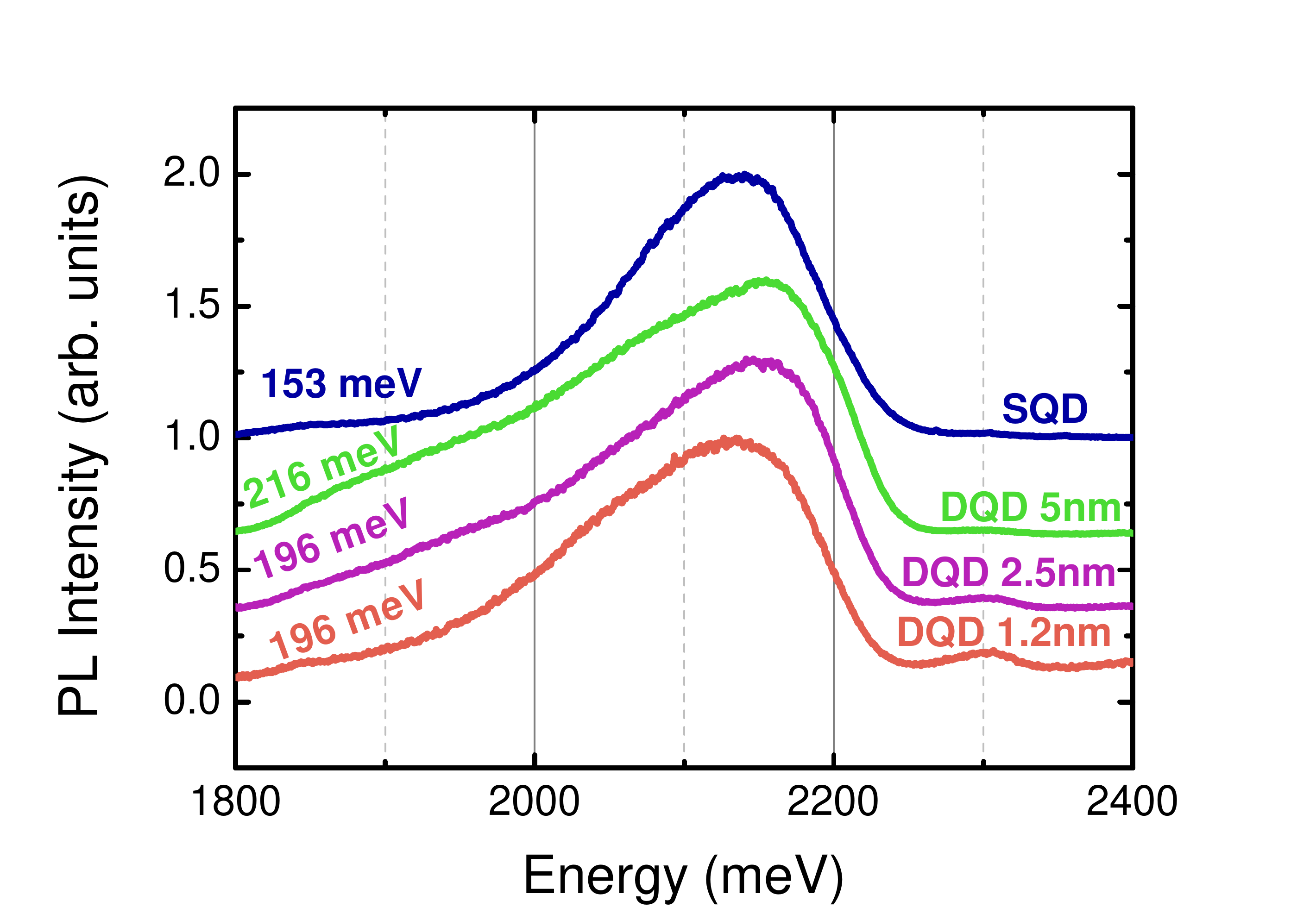}
  \caption{PL spectra of QD ensembles comparing double quantum dot (DQD) S1 samples with a reference single quantum dot (SQD) sample. ZnTe spacer layer widths are given in the annotations on the right. FWHM of the spectra are given in the annotations on the left.}
  \label{plm1}
\end{figure}

More insight into the detuning of the states in adjacent dots can be obtained from PL studies on QD ensembles. In Fig. \ref{plm1}, we show normalized PL spectra for three double quantum dot (DQD) samples from the S1 series with different widths of the spacer layer and compare them with a single dot (SQD) sample. The most striking result is that the full width at half maximum (FWHM) of the DQD ensemble PL spectra is in the range between 190 meV and 210 meV, i.e., about 40\% larger than for the SQD sample with FWHM of about 150 meV. Moreover, the DQD PL clearly exhibits a low energy tail, absent in the spectrum of the SQD sample. The presence of this tail is consistent with our conclusions from the STEM studies. While the bottom layer of dots is grown in the same conditions as those in the SQD sample, the top layer has a considerably deeper confinement potential owing to a larger Cd content. Thus, roughly, the tail can be interpreted as the PL of the top QDs, while the main part of the spectrum which generally coincides with that of a SQD sample can be ascribed to the recombination from the bottom dot.

Small peaks seen around 2300 meV for the DQD sample with thinnest spacer layer are related to the ZnTe barrier layer. Their visibility reflects the fact that the total PL intensity decreases with spacer width. We interpret this effect as a progressive deterioration of the crystal quality of the nanostructures, when the bottom QD is covered with a ZnTe layer comparable to the dot height. Moreover, for the thinnest spacer sample, the low energy PL tail is much less pronounced than for thicker spacer samples. We thus conclude that when the spacer becomes thin enough, the QD growth begins to resemble that of a single layer. 

Further proof for large energy detunings between the states in adjacent dots grown in a ZnTe matrix is found in Stark spectroscopy: bias-dependent PL studies. Namely, we search for the signatures of coherent coupling by studying S1 structures in {\em p-i-}Schottky diodes, analogous to the ones we previously employed to study the quantum confined Stark effect of single QDs.\cite{klo11} This diode structure allows to apply an electric field pointing anti-parallel to the growth axis (negative electric field). The choice of the {\em p-i-}Schottky structure is made since for a structure in which the top dot exhibits a deeper confinement, negative electric field allows to tune the hole states to resonance.\cite{bra06} For S1 samples, in a survey of a few hundred QDs we do not observe any anticrossings.

We interpret this absence of molecular-like coupling as resulting from the difference in the morphologies of the top and bottom QDs. This difference clearly translates into an energy detuning between the states in adjacent QDs -- larger than the attainable tunability with electric field. We assume that the hole ionization energy (energy level position with respect to the barrier band edge) in the bottom QD is the same as for a dot in a single layer -- about 50 meV \cite{klo14}. For the top QD, we can estimate the ionization energy from the STEM studies. Since both the band offset and strain-induced band shifts scale linearly with the amount of cadmium in a QD, we estimate that the ionization energy for holes in the top QD is roughly 2.5 times larger, i.e., about 125 meV. Since the magnitude of accessible electric fields in our {\em p-i-}Schottky diodes is roughly 100 kV/cm\cite{klo11} and the distance between the adjacent two QDs is 5 nm, the relative energy shift of the hole states is about 50 meV. This is only 75\% of the average hole detuning between the adjacent QDs, which explains the absence of signatures of resonant hole couplings in our Stark spectroscopy measurements.

We solve the problem of the large discrepancy in carrier energies in adjacent QDs through engineering of the confinement of the top QD. For QDs grown via the Stranski-Krastanow procedure, the QD size depends on the amount of accumulated elastic energy in the strained CdTe layer. More elastic energy allows to form more facets and thus form smaller QDs. This effect is demonstrated, e.g., by a strong PL red-shift from CdTe dots formed on Zn$_{0.7}$Mg$_{0.3}$Te -- the dots are larger than those grown on ZnTe owing to a smaller lattice mismatch between CdTe and Zn$_{0.7}$Mg$_{0.3}$Te.\cite{tin04} Moreover, the inter-shell splittings, proportional to the strength of the lateral confinement, increase with the number of CdTe monolayers from which the dots are formed and thus with the amount of accumulated strain.\cite{kuk11}

In order to bring the carrier energies in adjacent QDs closer to resonance, we need to decrease the hole ionization energy in the top dot, e.g., by decreasing the amount of cadmium in the top dot with respect to the bottom one or by inducing a growth of a relatively smaller top QD. In the S1 samples, after the bottom layer of dots is formed, the subsequent barrier layer is subject to tensile strain as the ZnTe lattice atoms tend to align with the dangling bonds of the CdTe QD. Thus, the ZnTe spacer layer effectively exhibits a {\em larger} lattice constant compared to the outer barrier, in effect reducing the compressive strain needed for QD formation. To counter this problem, in S2 samples, we grow the QDs with outer barriers made of \ZnMgTe, while the inner spacer layer is ZnTe. Since the lattice constant of ZnTe is smaller than that of the \ZnMgTe, the ZnTe spacer layer assures effectively {\em smaller} lattice constant then if this layer was grown, as the barrier, from \ZnMgTe. Consequently, the ZnTe spacer provides more elastic energy available for the formation of the top QD.

\begin{figure}[!h]
  \includegraphics[angle=0,width=.50\textwidth]{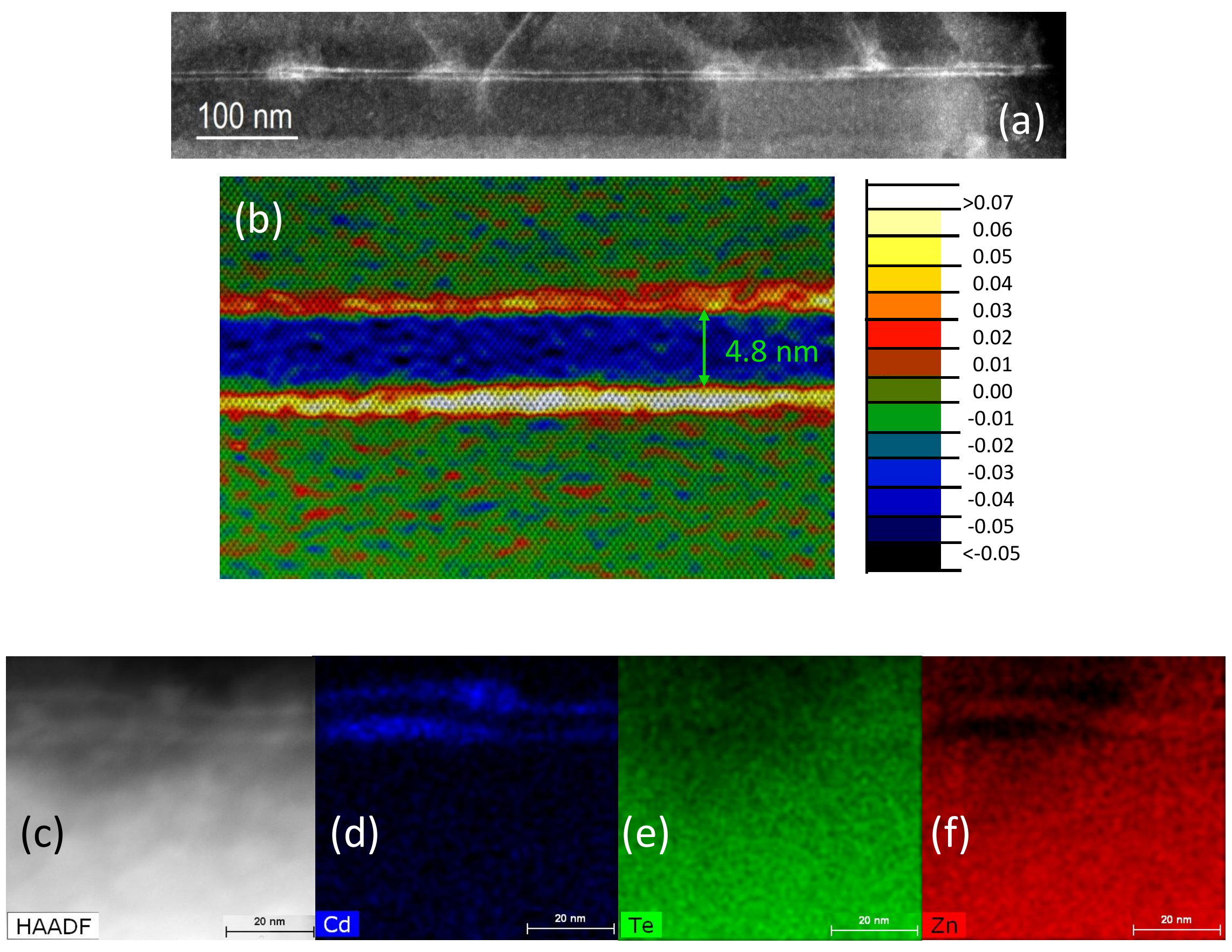}
  \caption{STEM analysis of a S2 sample with a 5 nm ZnTe spacer. (a) HAADF image showing two layers of QDs. (b) Atomic resolution lattice distortion map revealing a higher incorporation of cadmium in the bottom QD layer. (c) STEM HAADF image of a pair of quantum dots with the corresponding EDX maps of (d) Cd; (e) Zn; (f) Te}
  \label{mgtem}
\end{figure}

In Fig. \ref{mgtem}(a), we show the HAADF image of a S2 sample with a ZnTe spacer thickness of 5 nm. Again, vertical correlations in QD positions are observed. The lattice distortion map of a smaller sample cross-section is presented in Fig. \ref{mgtem}(b). The reference level is now the \ZnMgTe\ lattice constant. Two layers of dots are clearly resolved. Contrary to the S1 samples, a region of negative $\varepsilon_{002}$ is now seen in between the dot layers. This results from the ZnTe lattice constant being smaller than that of the surrounding \ZnMgTe\ matrix. It also shows that in this case the spacer layer accumulates elastic energy, which in turn results in the height of top QD smaller than that of the bottom QD. Also, contrary to the case of S1 samples, the lattice distortion map shows that the top QD contains less cadmium than the bottom one. The average lattice distortion parameters for the bottom and top QDs are 0.06 and 0.04, respectively. Thus, they differ by about 50\% as opposed to a factor of 2.5 in the case of S1 samples. Moreover, the absolute values of $\varepsilon_{002}$ for QD layers are larger than for S1 samples pointing to a larger cadmium content in S2 samples. Furthermore, we expect that in general the QD is S2 samples are larger than those in S1 samples owing to the smaller lattice mismatch between CdTe and \ZnMgTe\ matrix. However, we cannot directly prove this point with TEM studies, since it would require morphology investigations of a larger number of QDs.


\begin{figure}[!h]
  \includegraphics[angle=0,width=.5\textwidth]{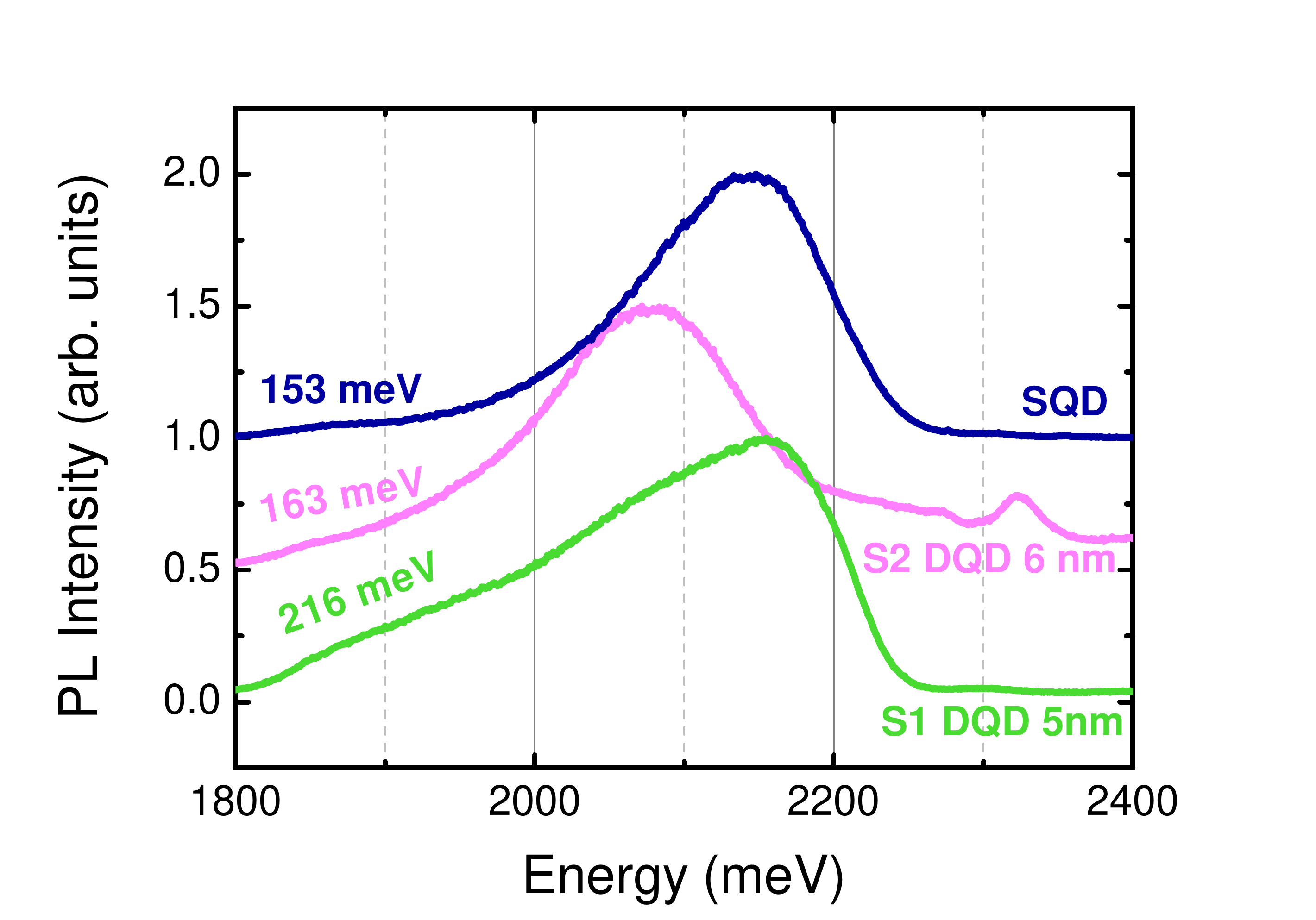}
  \caption{PL spectra of QD ensembles comparing double quantum dot (DQD) S1 sample with a reference single quantum dot (SQD) sample and a DQD S2 sample. ZnTe spacer layer widths are given in the annotations on the right. FWHM of the spectra are given in the annotations on the left.}
  \label{plm2}
\end{figure}

Results of the Fig. \ref{mgtem}(b) essentially prove our concept of tailoring the QD morphology with strain compensation. In order to verify whether it also results in a system having smaller detunings between the adjacent dots, we first examine PL spectra of QD ensembles. In Fig. \ref{plm2}, we compare the PL spectrum of a S2 sample with previously studied SQD and DQD samples. It exhibits a FWHM much smaller than that of the S1 DQD samples -- only about 160 meV. Also, the PL line shape is rather symmetric with the low energy tail missing. As a consequence of a larger incorporation of cadmium and possibly a larger size, the PL spectrum is redshifted with respect to the S1 samples.

\begin{figure}[!h]
  \includegraphics[angle=0,width=.50\textwidth]{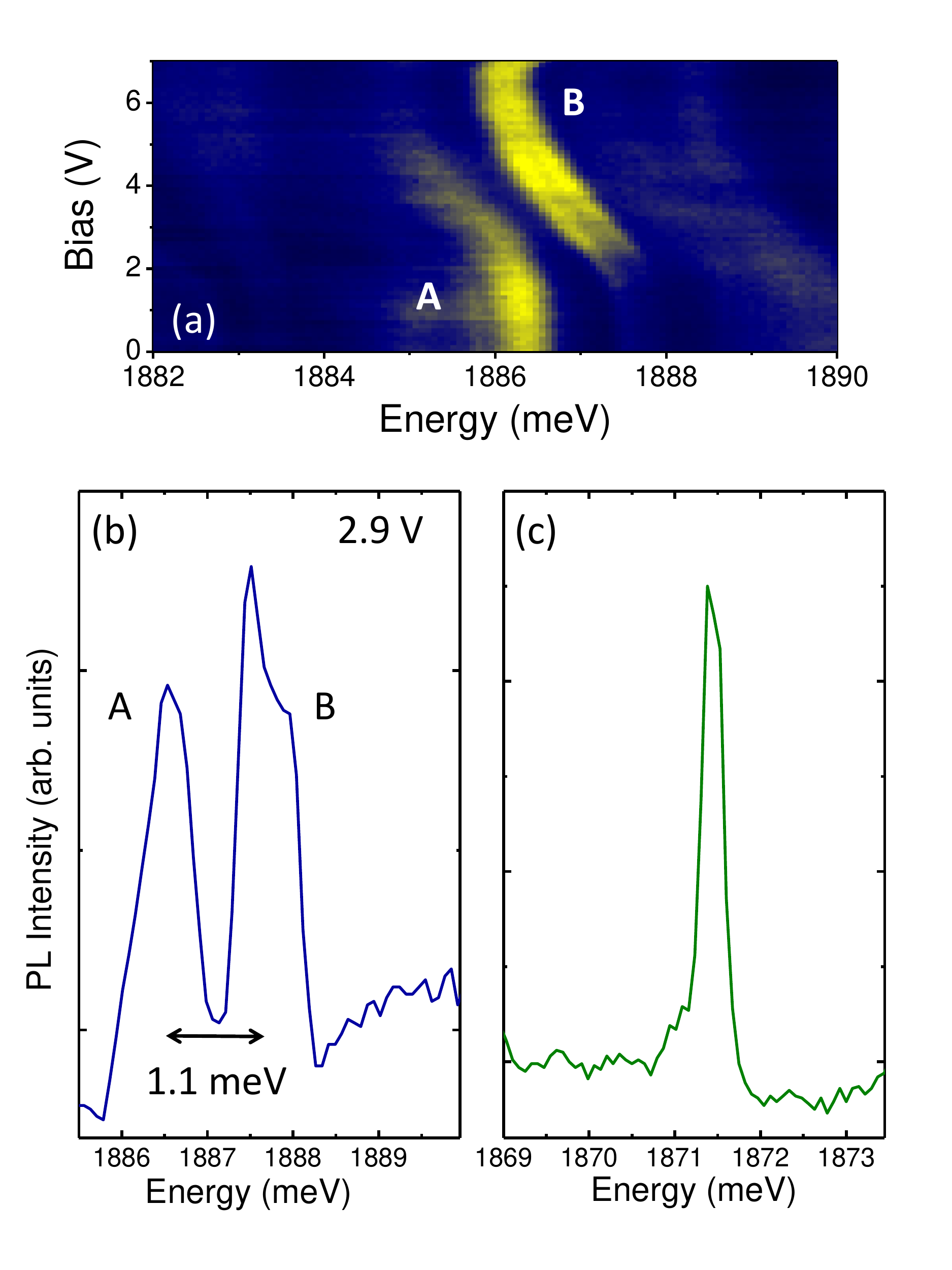}
  \caption{Start spectroscopy of a CdTe-based quantum dot molecule. (a) Map showing the PL dependence on the applied reverse bias. (b) PL spectrum at the resonant bias of 2.9 V. (c) Neutral exciton PL spectrum of from a single, uncoupled QD from the same bias scan.}
  \label{pl-map}
\end{figure}

The above results suggest a much smaller energy detuning between states in adjacent QDs. We now turn to Stark spectroscopy of single QDs in order to verify whether our growth procedure indeed results in QDs, where molecular coupling can be obtained. Since we aim at obtaining a resonance of hole states, for a sample with the bottom layer exhibiting a larger confinement we need to apply a positive electric field.\cite{bra06} Thus, we grow the S2 structures into an intrinsic region of a {\em n-i-p} diode, similar to the ones we used for Stark spectroscopy studies of \CdMnTe\ QDs.\cite{klo14} A map showing the dependence of the PL on the applied bias for a single QDM from S2 sample, with a ZnTe spacer width of 6 nm, is shown in Fig. \ref{pl-map}(a). With increasing reverse bias, transition A observed at 0 V redshifts by about 1 meV and eventually disappears at about 5 V. At about 1 V, transition B appears at higher energy and with increasing the reverse bias to 6 V redshifts also by about 1 meV. The PL spectrum at the resonance bias of 2.9 V is shown in Fig. \ref{pl-map}(b). It shows that the A and B transitions are separated by 1.1 meV. 
Transition linewidths are substantially larger than for single CdTe QDs\cite{bes01}. Indeed, a neutral exciton PL spectrum of an uncoupled QD from the same bias scan is shown in Fig. \ref{pl-map}(c).

\begin{figure}[!h]
  \includegraphics[angle=0,width=.50\textwidth]{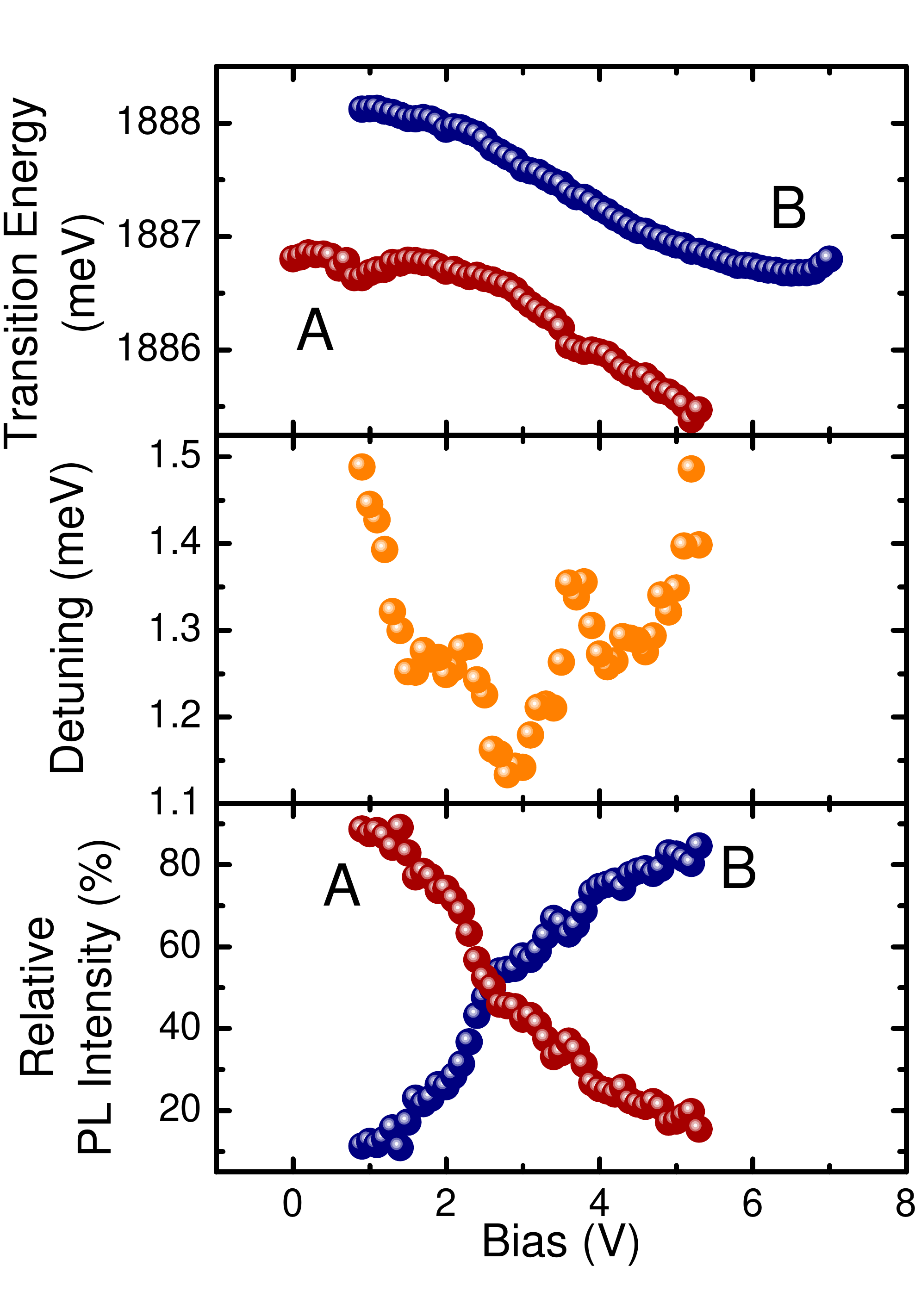}
  \caption{(a) Transition energy dependence on applied reverse bias for the A and B transitions shown in Fig. \ref{pl-map}(a). (b) Energy detuning between transitions A and B. (c) Relative PL intensity of the transitions A and B.}
  \label{pl-data}
\end{figure}

The anticrossing pattern has been reported by many authors as the fingerprint of the coherent coupling between intradot and interdot neutral excitons. The intradot exciton exhibits a small Stark shift stemming from a small built-in dipole moment on the order of 1 nm and electron-hole polarizability.\cite{fin04,klo11} On the contrary, the electron and hole in the interdot exciton are separated by the width of the spacer layer. Thus, the Stark shifts of the interdot excitons are several times larger.\cite{boy11jap} At resonance, the single carrier states hybridize over the two QDs, forming a bonding and an anti-bonding state. We interpret the anticrossing  shown in Fig. \ref{pl-map}(a) is an indication of the coherent coupling via hole tunneling in this QDM. The supporting arguments are shown in Fig. \ref{pl-data}, where we plot as a function of bias voltage the transition energies (a), the energy detuning between the two transitions (b), and their relative intensities (c) extracted by fitting two Gaussians. The detuning dependence on bias shown in Fig. \ref{pl-data}(b) allows to determine the resonance bias of 2.9 V. For larger (smaller) bias values, the ground state of the system is the interdot (intradot) exciton. However, in almost the whole bias range between 1 and 5 V, both states are mixed. This is seen in Fig. \ref{pl-data}(a): the bias dependence of the A transition flattens only close to 1 V. Thus, both states contain an admixture of the interdot exciton exhibiting a large dipole moment and being more sensitive not only to the external electric field, but also to local field fluctuations. These fluctuations result in the significant broadening of the PL transitions close to resonance as seen by comparing the spectra in Figs \ref{pl-map}(b) and (c). Finally, as shown in Fig. \ref{pl-data}(c), the intensity of the A decreases, while the intensity of the B transition increases with the bias. The exchange of oscillator strengths occurs at the resonance bias, which further supports our interpretation of the anticrossing as the resonance between inter- and intradot excitons. We also remark that the energy distance at the anticrossing is roughly equal to the values reported for InAs QDM.\cite{boy11jap}

The final proof for the interpretation of the anticrossing seen in Fig. \ref{pl-map}(a) as resulting from a coherent coupling would be a comparison of the Stark shift of the interdot exciton with the expected rate of $\sim F \cdot d$, where $F$ is the electric field and $d$ the width of the spacer layer. In the investigated {\em n-i-p} diodes however, the evaluation of the electric field is hindered by a strong leakage current present under reverse bias and originating probably from a large number of dislocations.\cite{klo14} This explanation is supported by negligible Stark shifts of uncoupled, intradot excitons observed in the same sample. Thus, a further optimization of these structures are needed to allow for quantitative analysis of the coupling effects.

\section{Conclusions}

In summary, we have shown that the method of strain compensation can be successfully applied to engineer the quantum confinement in CdTe quantum dot molecules. Specifically, growing a spacer layer made of a material with a smaller lattice constant than the outer barriers counteracts the decrease of available elastic energy for the formation of the top dot layer. As a consequence, it is possible to obtain a smaller top quantum dot and energetically less detuned from the bottom one. The magnitude of the detuning can be monitored by evaluating the FWHM of the ensemble quantum dot photoluminescence. For confinement-engineered structures, the FWHM are less than 10\% larger than for single dot samples. On the other hand, structures grown with uniform ZnTe barriers exhibit ensemble PL broadening of about 40\%.  Proper engineering of the confinement in dots embedded in diode structures allows for tuning of the hole state to resonance. This results in observation of anticrossings in bias dependent photoluminescence studies, pointing to formation of a molecular-like state, hybridized over the coupled dots. Our results pave the way towards advanced studies of II-VI quantum dot molecules, such as coupling of semimagnetic dots, dots containing single magnetic ions, or structures based on selenides.

\begin{acknowledgments}
This research was supported by the Polish National Science Center grant no. 2011/01/B/ST3/02287. T.W. acknowledges the support from the Foundation for Polish Science through the International Outgoing Scholarship 2014. Part of the The TEM investigations was supported by European Regional Development Fund through the Innovative Economy grant (POIG.02.01-00-14-032/08).
\end{acknowledgments}

\end{document}